\documentclass[article, twocolumn]{revtex4}
\usepackage{graphicx}
\usepackage{color}
\usepackage{amsmath}

\newcommand{\ket}[1]{\left| #1 \right\rangle}
\newcommand{\bra}[1]{\left\langle #1 \right|}

\begin{document}
\title{Verifying entanglement in the Hong-Ou-Mandel dip}
\author{Megan R. Ray and S. J. van Enk}
\affiliation{Department of Physics\\
Oregon Center for Optics\\University of Oregon, Eugene OR 97403}

\begin{abstract}
The Hong-Ou-Mandel interference dip is caused by an entangled state, a delocalized bi-photon state. We propose a method of detecting this entanglement by utilizing inverse Hong-Ou-Mandel interference, while taking into account vacuum and multi-photon contaminations, phase noise, and other imperfections. The method uses just linear optics and photodetectors, and for single-mode photodetectors we  find a lower bound on the amount of entanglement.  
\end{abstract}

\maketitle
\section{Introduction}

Quantum interference effects that arise when single photons impinge on a beam splitter are crucial to linear-optics quantum computing schemes  \cite{O'Brien,KLM,Kok}, with the other indispensable {\em non}linear ingredient provided by photon-counting measurements. 
One such linear-optics quantum interference effect was observed for the first time in 1987, by Hong, Ou, and Mandel, and it still carries their name \cite{HOM}. In the Hong-Ou-Mandel interference (HOMi) effect, two photons in otherwise identical modes impinge on two different input ports of a 50/50 beam splitter, and, thanks to bosonic interference, always emerge together in one of the two output ports. More precisely, the output state can be expressed in Fock states as
\begin{equation}\label{HOMstate}
\ket{\Psi}_{AB}=(\ket{0}_A\ket{2}_B-\ket{2}_A\ket{0}_B)/\sqrt{2}.
\end{equation}
Here $A$ and $B$ denote the two output modes, with identical polarizations, frequencies, and transverse spatial quantum numbers, and differing only in their propagation directions.
Great progress has been made recently in building waveguide circuits on chips, with which high-visibility interference fringes involving multi-photon states with high purity such as $\ket{\Psi}$  can be observed \cite{OBrienNP}.

The aspect of the output state $\ket{\Psi}_{AB}$ that interests us here is that it, provided the modes $A$ and $B$ are spatially separated, is entangled.
For instance, the pure state $\ket{\Psi}$ can be shown to violate Bell-type inequalities \cite{Wildfeuer}. What concerns us in particular, is how one could verify the entanglement of noisy versions of the ideal state, containing, e.g., phase noise and contaminations with states with different numbers of photons (no photons at all, one photon in total, or more than two photons in total).
As it turns out, standard measurements and operations used in, e.g.,  \cite{OBrienNP} to characterize and manipulate few-photon states are indeed sufficient for entanglement verification, provided (but this is a far from trivial proviso) all photo detectors detect photons only in particular modes. That is,
if we assume our detectors are sensitive only to one particular polarization, spectral profile, and transverse spatial mode, then the method we present here will  unambiguously detect entanglement even if the actual input state (with arbitrary numbers of photons in it) has a multi-mode character.
 Moreover, in this case we can construct lower bounds on the amount of entanglement as well. 
 The reason is, that such a detection scheme is equivalent to
 a protocol where a filtering operation is applied to the input state that keeps only photons in  the desired modes. Since this operation is {\em local}, the amount of entanglement of the resulting filtered state cannot be larger, on average, than the entanglement present in the input state.
 
On the other hand, if we drop the assumption about the single-mode character of our detection devices, then the problem of verifying entanglement of a delocalized two-photon state becomes much more involved, also when compared to the similar problem of verifying entanglement of a delocalized {\em single} photon \cite{Papp,Pavel}. We will give  the essential reason for this difference and present solutions for the multi-mode multi-photon entanglement verification problem that will work if the state under investigation is sufficiently close to a single-mode entangled state.

It may be interesting to compare our entanglement verification scheme to a scheme proposed in Refs.~\cite{Loss, Fazio}, which likewise uses the HOM interference effect (but in its fermionic version) to detect entanglement. The latter scheme detects entanglement between electrons, and assumes the number of electrons in each input port of a 50/50 beamsplitter is fixed and known, whereas we do not assume a fixed photon number. Indeed, such an assumption is perfectly fine for first-quantized electrons, but not for second-quantized photons. Moreover,
we use the {\em inverse} HOM effect to detect entanglement in a state: ideally, we have either two photons  or no photons in each input mode, whereas Refs.~\cite{Loss, Fazio} consider, in the ideal case, one electron in each input mode, and then use the proper HOM effect for entanglement detection. 

Finally, we recall that the (proper) HOM effect has been used to detect entanglement between two input photons (see, e.g., Ref.~\cite{Giovannetti} and references therein). It's still true that the assumption that there is exactly one photon in each input port is not warranted in general, but, for entanglement verification, it is an allowed filtering operation, as it is local. In contrast, filtering on having two photons in total in the two input ports (which operation we would like to perform for our case) would be nonlocal. Also note that in our case, the output of the inverse HOMi experiment would ideally be a product state of two photons.

\section{Entanglement verification for single-mode states}
\subsection{Defining ``single mode''}
Let us first consider so-called single mode states, by which we mean states where any photons present are in the same transverse spatial, spectral, and polarization modes, with the understanding that they can differ in their direction of propagation (there are two such modes in our case, spatially separated, which we call modes A and B). 
Since experiments typically must be repeated in time, we do allow the spectral mode functions $\phi(\omega)$ to differ by a phase factor $\exp(i\omega T)$ with $T$ a {\em known} delay time, without the photons losing their single-mode character. 

We could, in principle, perform tomography on the full state to determine its density matrix and from this calculate a measure of entanglement, e.g., the concurrence or negativity of the state, and thus determine whether the state is entangled. However, since we shouldn't  assume anything about the Hilbert space that the state lives in (since we want be able to verify the entanglement on noisy versions of our ideal state), we would have an infinite number of matrix elements to determine.  Even if we were to make restrictive assumptions about the Hilbert space of the state, it would still require numerous measurements to fully determine the state.  For example, if we assumed that the state did not contain more than two photons, this would still leave a 6x6 density matrix to determine. If we are not interested in fully characterizing the state, but merely in verifying its entanglement we do not need to do so much work.  Instead of trying to exactly calculate a measure of entanglement of the state, we can instead calculate a lower bound  which will allow verification of entanglement of the state with far fewer measurements.   
\subsection{Local filtering}
Let the state whose entanglement we are trying to verify be called $\rho$. A bound on the entanglement can be found in the following way.  Suppose we were to apply the following {\em local} filtering operations: we ask about each of the two spatially separated modes $A$ and $B$ two questions 
\begin{align}
&{\bf Filter\, ``1'':}\,\, \textrm{Is there exactly 1 photon in the mode?}\nonumber\\
&{\bf Filter\, ``2'':}\,\,\textrm{Are there more than 2 photons in the mode?}\nonumber
\end{align}
We consider this filtering a success if the answer is ``no'' to both questions [cf. Eq.~(\ref{HOMstate})]. The probability then of successful filtering is $\tilde{P}=P_{0,0}+P_{0,2}+P_{2,0}+ P_{2,2}$, where $P_{i,j}$ is the probability to find $i$ photons in mode A and $j$ photons in mode B in the unfiltered state $\rho$. This filtering collapses our state to one living in the smaller Hilbert space spanned by $|0\rangle_{A}|0\rangle_{B}$, $|0\rangle_{A}|2\rangle_{B}$, $|2\rangle_{A}|0\rangle_{B}$, and $|2\rangle_{A}|2\rangle_{B}$. At this point we have a state represented by a density matrix with up to 16 nonzero elements. To simplify calculations we can further bound the state's entanglement by assuming we apply another local operation, which in addition requires classical communication: 
\begin{eqnarray*}
&{\bf Local\,\, operation+CC:\,\, ``phase shift'':}\\
& \textrm{apply the same random phase shift to both modes}
\end{eqnarray*}
thus destroying any coherence between states with different numbers of photons and reducing the number of nonzero matrix elements to at most 6.
\subsection{Entanglement criterion}    The end result of filtering is of the  (normalized) form
\begin{equation}
\tilde{\rho}=\frac{1}{\tilde{P}}\left(\begin{array}{cccc}P_0 & 0 & 0 & 0 \\0 & P_{0,2} & d & 0 \\0 & d^* &  P_{2,0} & 0 \\0 & 0 & 0 & P_{2,2}\end{array}\right)
\end{equation}
Since concurrence is an entanglement monotone and $\tilde{\rho}$ is the result of only local operations and classical communication applied to $\rho$, the concurrence of $\tilde{\rho}$ bounds the concurrence of $\rho$: $ \tilde{P} C(\tilde{\rho})  \leq C(\rho)$. The concurrence of $\tilde{\rho}$ is 
\begin{equation}
\tilde{P} C(\tilde{\rho}) = \max[\:0,\:2|d| -2\sqrt{P_{0}P_{2,2}}\: ]
\end{equation}
which is greater than zero when
\begin{equation}\label{ineq1}
P_{0}P_{2,2}<|d|^2.
\end{equation}
Thus $\tilde\rho$ is provably entangled if ineq.~(\ref{ineq1}) holds true, and so too is $\rho$. 
Similarly, since negativity is also an entanglement monotone, the negativity of $\tilde{\rho}$ bounds the negativity of $\rho$ in the same way: $\tilde{P} \mathcal{N}(\tilde{\rho}) \leq \mathcal{N}(\rho)$. But calculating the negativity of $\tilde{\rho}$ results in exactly the same bound as found by calculating the concurrence: the state is provably entangled if $P_{0}P_{2,2}<|d|^2$.  

Now we must find a way to bound $|d|^2$.  Since  $d=\tilde{P}\langle02|\tilde{\rho}|20\rangle=\langle02|\rho|20\rangle$ we don't need to physically perform any of the filtering mentioned above, as we can determine the needed information, $d$, from the unfiltered state $\rho$.  To do this, consider placing the two modes of $\rho$ on the two input ports of a lossless 50/50 beamsplitter.  We will label the input modes A and B, and the output modes C and D.  The transformation between input mode creation operators and output creation operators can be written as follows (after adding, for convenience, a $\pi/2$ phase shift to mode D to compensate for the $\pi/2$ phase shift upon reflection)
\begin{equation}
a^{\dag} \to \frac{c^{\dag} + d^{\dag}}{\sqrt{2}} \qquad \text{and} \qquad b^{\dag} \to \frac{c^{\dag} - d^{\dag}}{\sqrt{2}}
\end{equation}
which allows us to calculate photo-detection probabilities $Q_{i,j}$ for the output modes, where $Q_{i,j}$ is the probability to find $i$ photons in mode C and $j$ photons in mode D.  It can be shown that 
\begin{equation}
Q_{1,1}=\frac{1}{2}\left(P_{2,0}+P_{0,2}-d-d^*\right),
\end{equation}
which gives
\begin{equation}
\left(Q_{1,1}-\frac{P_{2,0}+P_{0,2}}{2}\right)^2= \left(\frac{d+d^*}{2}\right)^2=\Re(d)^2\leq |d|^2.
\end{equation}
So when 
\begin{equation}
\label{identical}
P_0P_{2,2}<\left(Q_{1,1}-\frac{P_{2,0}+P_{0,2}}{2}\right)^2
 \end{equation}
the state can be said to be provably entangled.  Figure 1 plots both sides of our inequality (\ref{identical}) for many randomly picked separable states, to show how this criterion indeed verifies entanglement. Moreover, the figure caption identifies the states lying on the borderline between separable and verifiably entangled. 
\begin{figure}
\label{combinedplot}
  \includegraphics[width=7cm]{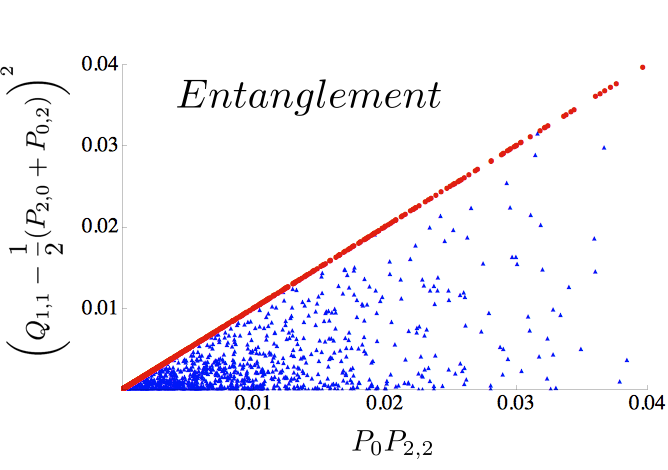}
  \caption{Scatter plot of
the right-hand side vs the left-hand side of our entanglement criterion (\ref{identical}). Red dots lie on the boundary of entanglement vs separable, and correspond  to
pure  separable states of the form $(|0\rangle_A+a|2\rangle_A)\otimes(|0\rangle_B+b|2\rangle_B)$ where $a$ and $b$ are real. Blue triangles corresponds to mixtures of two randomly generated separable states of the form $(|0\rangle_A+a_1|1\rangle_A+a_2|2\rangle_A)\otimes(|0\rangle_B+b_1|1\rangle_B+b_2|2\rangle_B)$ (with complex coefficients).  }
\end{figure}
\subsection{An additional phase shift}
Our condition (\ref{identical}) will not detect entanglement in an input state, even when it is in fact present, when $d$ is largely or purely imaginary. But if one were to apply a phase shift to one of the modes before placing the state on the beam splitter and vary that phase until $Q_{1,1}$ was maximized (the same local operation with classical communication as performed in  \cite{OBrienNP}), this would maximize $\Re(d)^2$, thus making ineq.~(\ref{ineq1}) equivalent to (\ref{identical}). In other words, such states then can be detected by our criterion. 
Take, for instance, the state 
\begin{equation}\label{rho1}
\rho_1:=\frac{1}{6}\ket{00}\bra{00}+\frac{1}{3}(\ket{20}+i\ket{02})(\bra{20}-i\bra{02})+\frac{1}{6}\ket{22}\bra{22}.
\end{equation}  For this state $|d|^2=\frac{1}{9}$ and $P_0P_{2,2}=\frac{1}{36}$ so by ineq. (\ref{ineq1}) the state is in fact entangled.  However $\Re(d)^2=0$, so ineq. (\ref{identical}) will not detect the entanglement. But if we apply a phase shift of $\exp(i\frac{\pi}{2})$ to one of the modes then $d$ will become purely real (and so $Q_{1,1}$ will be maximized), and ineq.~(\ref{identical}) {\em will} detect the entanglement. As Figure 2 (top) shows, for this state with a phase $\exp(i\phi)$ applied to the first mode, entanglement will be detected when $\phi$ is between $\frac{1}{6}\pi$ and $\frac{5}{6}\pi$ or between $\frac{7}{6}\pi$ and $\frac{11}{6}\pi$.  A similar, but more noisy state, 
\begin{equation}\label{rho2}\rho_2:=\frac{1}{3}\ket{00}\bra{00}+\frac{1}{4}(\ket{20}+i\ket{02})(\bra{20}-i\bra{02})+\frac{1}{6}\ket{22}\bra{22},\end{equation} 
will have a smaller range of detectable entanglement, specifically when $\phi$ is between $.39\pi$ and $.61\pi$ or between $1.39\pi$ and $1.61\pi$ (see Figure 2, bottom part).

\begin{figure}[h]
\begin{center}$
\begin{array}{cc}
\includegraphics[width=7cm]{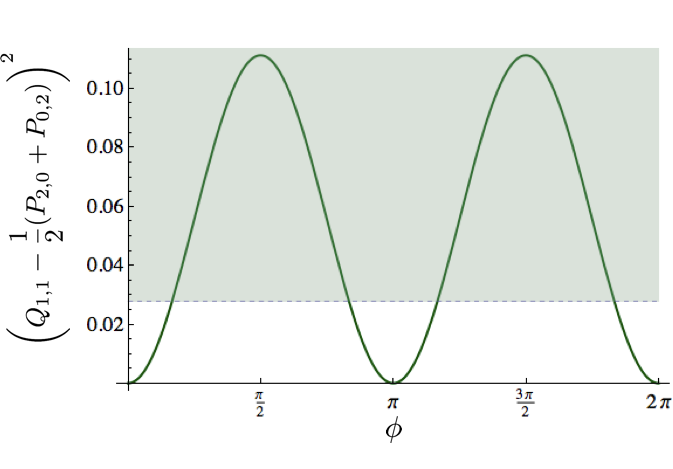} \\ 
\includegraphics[width=7cm]{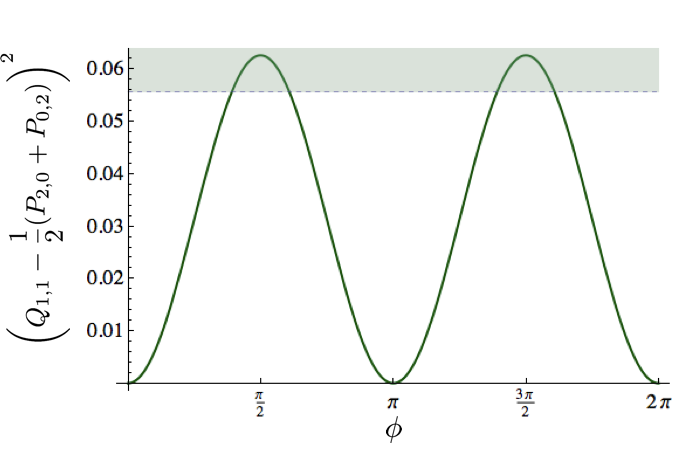}
\end{array}$
\end{center}
  \caption{$\left(Q_{1,1}-\frac{1}{2}(P_{2,0}+P_{0,2})\right)^2$, that is, the right-hand side of inequality (\ref{identical}), for the state $\rho_1$ (top), defined in (\ref{rho1}), and the more noisy $\rho_2$ (bottom), defined in (\ref{rho2}), as a function of a phase shift $\exp(i\phi)$ applied to the first mode. The shaded region represents for which values of $\phi$ entanglement will be detected by ineq.~(\ref{identical}) [both states are entangled for any value of $\phi$].}
\end{figure}
\subsection{Asymmetric beamsplitters}
To bound $d$ we placed our state on a 50/50 beamsplitter, but it is easy to generalize our analysis to beam spitters which are not equally balanced.  Suppose our beam splitter has a (real) reflection coefficient $r$ and a (real) transmission coefficient $t=\sqrt{1-r^2}$ such that  the input creation operators  transform as
\begin{equation}
a^{\dag} \to rc^{\dag} + td^{\dag} \qquad \text{and} \qquad b^{\dag} \to tc^{\dag} - rd^{\dag}
\end{equation}
Following the same analysis as before we find that if
\begin{equation}
P_0P_{2,2}<\left(\frac{Q_{1,1}+P_{1,1}(t^2-r^2)}{4r^2t^2}-\frac{P_{2,0}+P_{0,2}}{2}\right)^2
\end{equation}
the state is provably entangled.

\section{Considerations concerning multi-mode multi-photon states}
We made the assumption at the beginning of our analysis that any photons present are in the same transverse spatial, spectral, and polarization mode.  However if our detectors only detect a certain single mode we can drop the assumption that the photons being in the same mode as this is equivalent to a local filtering. That is, using single mode detectors is equivalent to an additional filtering performed on each of the spatially separated modes, filtering out all photons not in the single mode of interest before detection takes place. What if we drop the assumption of single-mode detectors?

Suppose we have an input state in which the photons present are {\em not} all in the same transverse spatial, spectral, and polarization mode.
The entanglement verification scheme described above did assume that the two photons in the filtered state (after the local filtering operations ``1'' and ``2'') are in the same mode, because of the explicit assumption that there is interference (of the ``inverse HOM'' type) taking place on a beam splitter. 
But this assumption does affect how we interpret the results of the measurements: in particular, the quantity $Q_{11}$ (which we would like to be large) could be dangerously contaminated with contributions from those input states that lead to larger values of $Q_{11}$ for photons in different modes than for photons in the same modes. For example, if we start with an output state with one photon in each output port, but of different colors, then applying the inverse beam-splitter transformation yields an input state that has this undesired property. The question is to what extent we can avoid or correct for the presence of such input states. 
\subsection{A corrected entanglement criterion}
One way of correcting for these unwanted states is to subtract the contribution from the worst possible kind of state, i.e., one that maximizes the right hand side of Eq.~(\ref{identical}) without HOM entanglement, such as the state mentioned above
\begin{equation}
\left(\ket{10}_{red} +\ket{01}_{red}\right)\otimes\left(\ket{10}_{blue} -\ket{01}_{blue}\right)/2
\end{equation}
While this state has twice as much entanglement as the HOM state, it is not the type of entanglement we are interested in trying to detect here.  A state such as this with a probability $P^o_2$ of detecting two photons of different color will contribute at most $3P^o_{1,1}/2$ to the quantity being squared on the rhs of Eq.~(\ref{identical}), so we will compensate for this possible contribution by subtracting $3P^o_{1,1}/2$.  For states close to the ideal state the contamination of different colored photons will be small and thus the correction will be small. We can also construct a bound that does not rely on measuring the probability of detecting two photons of different colors, since it is always less than or equal to the probability of detecting two photons of any color($P^o_{1,1}\le P_{1,1}$.) Using this, our (conservative) condition for entanglement becomes
\begin{equation}\label{conservative}
P_0P_{2,2}<(\max[\: Q_{1,1} - P_{1,1} -P_2/2\:,0\:])^2
\end{equation}
\subsection{Nonexistence of local  filters for sameness of modes}

It would be nice if we could find a local filtering operation that checks whether two input photons propagating in one direction are in the same mode with respect to the other quantum numbers or not. There is certainly no von Neumann measurement that achieves that goal, as the target states are not all orthogonal. But, surprisingly, we cannot even construct a POVM that does the trick: the reason is that even if we start with a state that contains two photons in orthogonal modes, say described by creation operators $a_1^\dagger$ and $a_2^\dagger$, then we can view the same state as a superposition of two states, each with the two photons in identical modes, as described by the creation operators
$a_{\pm}^\dagger=(a_1^\dagger\pm a_2^\dagger)/\sqrt{2}$. This results from the identity
\begin{equation}
a_1^\dagger a_2^\dagger=\frac{(a_+^\dagger)^2
-(a_-^\dagger)^2}{2}.
\end{equation}
This is then the essential difference between single-photon states and multi-photon states, which makes entanglement verification much harder for two-photon states than for single-photon states!
Moreover, this also illustrates a difference between bosons and fermions: in the case of two fermions there {\em is} an antisymmetric subspace, and, e.g., we can certainly perform a measurement that checks whether two spin-1/2 systems have different spins (singlet state!) or not.
\subsection{An alternative local operation}
All is not quite lost, as we can still apply other sorts of local operations that are useful for the analysis of entanglement of the input state. In particular, suppose that our input state is some coherent superposition of, e.g., 
the desired state $\ket{0}_A\ket{2}_B-\ket{2}_A\ket{0}_B$ and an unwanted state $\ket{1}_{A_1}\ket{1}_{A_2}\ket{0}_B$ (with photons in different modes).
There is a local operation that transforms this superposition into an incoherent mixture of these two states: for each pair of orthogonal modes $A_k$ and $B_k$ (picked from some fixed basis: that's the essential difference from the no-go statement from the preceding subsection) apply a random $k$-dependent phase shift, and then forget the precise phase shifts applied. This operation
will only preserve the coherence of
superpositions of photons in the same spectral, polarization and transverse modes in $A$ and $B$. That is, by a local operation we can transform the input state into a state of the form
\begin{equation}
\rho=P_s\rho^s+(1-P_s)\rho^{\perp},
\end{equation}
where the first term denotes states that do display (inverse) HOM interference, and the second term states that do not; $P_s$ is the 
probability of observing HOM interference, given $\rho$.
The point is that we have now separated the input state in two parts, the first part of which is the state for which our method demonstrates entanglement (see below for further elaborations of this point). The second term has no entanglement, since any superpositions in that term have been destroyed. Its presence could imply the state $\rho$ is not entangled, even if $\rho^s$ is, namely if $1-P_s$ is too large. We will not solve the (hard) general problem of identifying for what values of $P_s$ and for what states $\rho^s$, entanglement of the latter still implies entanglement of $\rho$.

Let us return to the statement that $\rho^s$ is entangled, if our verification method succeeded.
We still have to discuss the fact that our method assumed that both photons are in one particular mode, whereas for photons in $\rho^s$ we only know they are in the same mode, but not in which one. This does have consequences for the amount of entanglement (see \cite{Jun} for extensive discussions of this issue), but not for the bare fact that the state is entangled. We can demonstrate this by showing that the state $\rho^s$ can be distilled (the following protocol is far from optimal, and one can easily improve its efficiency; here its point is only an existence proof): just take two copies of $\rho^s$; first determine a particular mode such that the projection of $\rho^s$ onto that mode is entangled;
then
perform on each of the $A$ and $B$ modes a joint measurement that counts how many photons in that particular mode there are in total in the two copies. If the answer is ``2'' for both $A$ and $B$, we have an entangled state in that one particular mode.
In this highly inefficient protocol the average amount of entanglement decreases (unless only a single mode is occupied), but it stays nonzero.
Hence $\rho^s$ must be entangled.

\section{Summary}
We demonstrated how the inverse HOMi effect can be used to verify the mode entanglement present in a state of the form $(\ket{0}\ket{2}-\ket{2}\ket{0})/\sqrt{2}$, and noisy versions of it.
If the photons in the state are all ``single-mode'', that is, all have the same polarization, the same transverse mode profile and the same spectral profile, then our method easily bounds the amount of entanglement from below. 
That directly gives a criterion, inequality (\ref{identical}), which, when satisfied for a given single-mode state, is sufficient to prove entanglement. We analyzed how the applicability of the criterion can be improved simply by applying an additional phase shift to one of the two modes. The operations needed to verify entanglement can be implemented with linear optics, and are just those demonstrated in the experiment of \cite{OBrienNP}.

We discussed how the problem of verifying entanglement in the delocalized two-photon state with the inverse HOMi effect becomes more ``interesting'' (a euphemism for ``complicated'') without this single-mode assumption [more precisely, when both the input state {\em and} one's photo detectors are multi-mode], and why a delocalized single-photon state does not suffer from these complications. On the other hand, the interpretation of violating a Bell inequality with unbalanced homodyne measurements \cite{Wildfeuer} is immune to the single-mode or multi-mode character of the input state, at the small cost of requiring  phase-locked local oscillators, thus showing an advantage of Bell inequalities in the context of entanglement verification.

We gave a simple solution to the full problem of inverse HOMi multi-mode multi-photon mode entanglement, based on bounding the deviation of the actual state  from a single-mode state, which works well when that deviation is sufficiently small.
It yields a (more conservative) entanglement criterion (\ref{conservative}).

\end{document}